\documentclass[prl,amsfonts,twocolumn]{revtex4}
\usepackage{graphicx, epsfig}
\usepackage{color}
\usepackage{mathrsfs}
\usepackage{amsmath}

\newcommand{\be}{\begin{equation}}
\newcommand{\ee}{\end{equation}}
\newcommand{\bea}{\begin{eqnarray}}
\newcommand{\eea}{\end{eqnarray}}

\newcommand{\gapp}{\mathrel{\raise.3ex\hbox{$>$}\mkern-14mu \lower0.6ex\hbox{$\sim$}}}
\newcommand{\lapp}{\mathrel{\raise.3ex\hbox{$<$}\mkern-14mu \lower0.6ex\hbox{$\sim$}}}
\def\bbox{{\,\lower0.9pt\vbox{\hrule \hbox{\vrule height 0.2 cm
\hskip 0.2 cm \vrule  height 0.2 cm}\hrule}\,}}
\newcommand{\beq}{\begin{equation}}
\newcommand{\eeq}{\end{equation}}

\newcommand{\Mpl}{M_{\rm Pl}}

\begin{document}
\title{Detecting Vanishing Dimensions Via Primordial Gravitational Wave Astronomy}
\author{Jonas Mureika$^1$, and Dejan Stojkovic$^2$}
\affiliation{$^1$Department of Physics, Loyola Marymount University, Los Angeles, CA~~90045}
\affiliation{$^2$Department of Physics,
SUNY at Buffalo, Buffalo, NY 14260-1500}


\begin{abstract}

\widetext
Lower-dimensionality at higher energies has manifold theoretical advantages as recently pointed out in \cite{dejan1}. Moreover, it appears that experimental evidence may already exists for it - a statistically significant planar alignment of events with energies higher than TeV has been observed in some earlier cosmic ray experiments. We propose a robust and independent test for this new paradigm.  Since (2+1)-dimensional spacetimes have no gravitational degrees of freedom, gravity waves cannot be produced in that epoch. This places a universal maximum frequency at which primordial waves can propagate, marked by the transition between dimensions.  We show that this cut-off frequency may be accessible to future gravitational wave detectors such as LISA.
\end{abstract}


\maketitle


There is growing theoretical evidence to suggest that the short-distance spatial dimensionality is {\it less} than the macroscopically-observed three.   Causal dynamical triangulations \cite{cdt} demonstrate that the four-dimensional spacetime can emerge from two-dimensional simplicial complexes.  It has also been shown that a non-commutative quantum spacetime with minimal length scale will exhibit the properties of a two-dimensional manifold  \cite{lmpn1}. Reducing the number of dimensions in the far UV limit offers a completely new approach to gauge couplings unification \cite{Shirkov:2010sh}. An argument for dimensional reduction at high energies based on the strong coupling expansion of the Wheeler-DeWitt equation was presented in \cite{Carlip:2009km}.  In a similar vein, the cascading DGP model \cite{cdgp} provides a mechanism for the emergence of an extra spatial dimension only at Hubble scales, in order to solve the cosmological constant problem. It was even argued that evidence of higher dimensionality at cosmological scales is already present in the current observational data \cite{Afshordi:2008rd}.

\begin{figure}[htb]
\vspace*{-0.1in}
\center{\scalebox{0.25}{\includegraphics{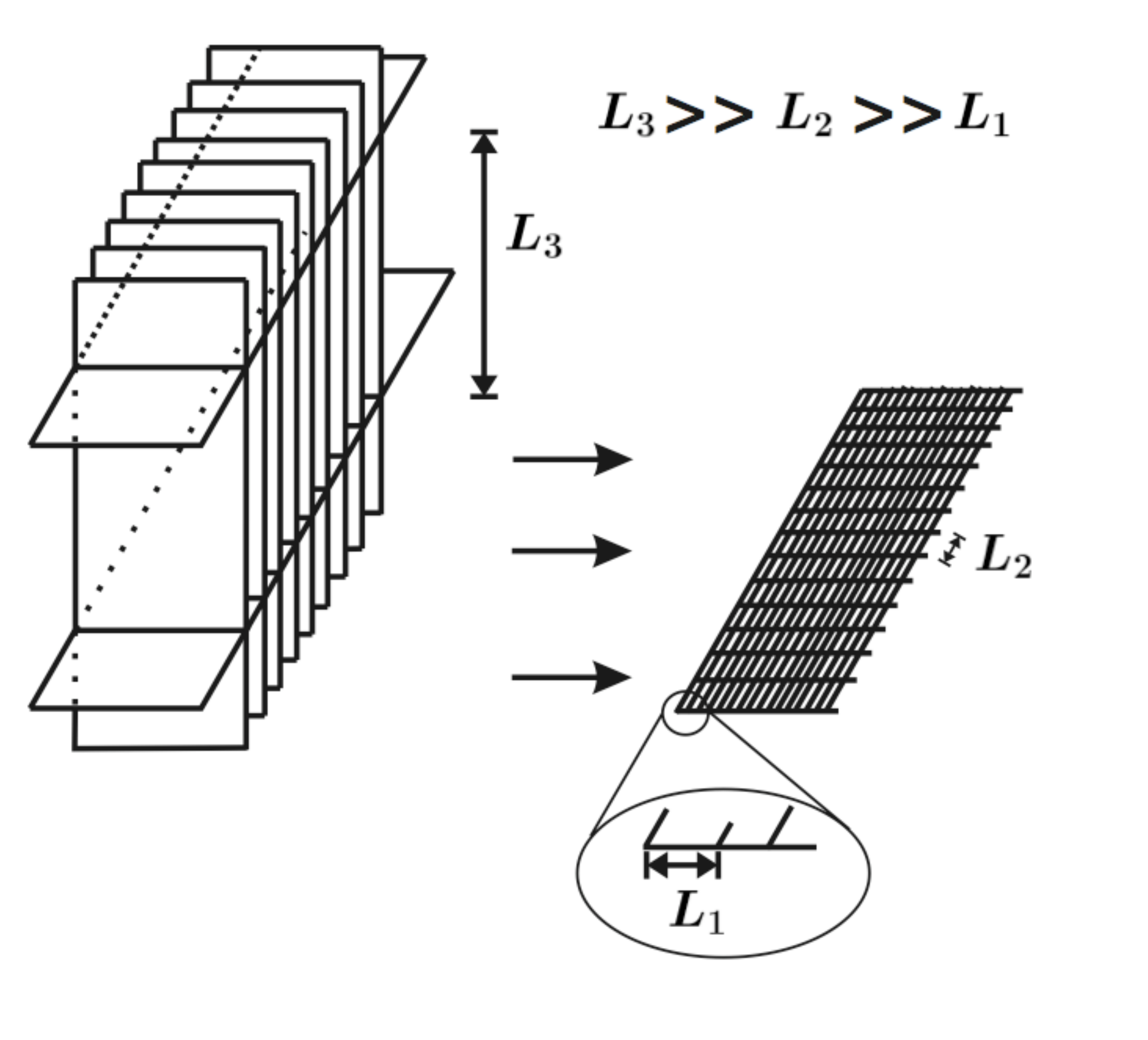}}}
\vspace*{-0.1in}
\caption{We assume that space-time has an ordered lattice structure, which becomes anisotropic at very small distances. The  fundamental quantization scale of space-time is indicated by $L_1$. Space structure is 1D on scales much shorter than $L_2$, while it appears effectively 2D on scales much larger than $L_2$ but much shorter than $L_3$. At scales much larger than
  $L_3$, the structure appears effectively 3D. Following this hierarchy, at even larger scales, say $L_4$, yet another dimension opens up and the structure appears 4D (not shown in the picture).}
\label{lattice}
\end{figure}

\begin{figure}[htb]
\vspace*{-0.1in}
\center{\scalebox{0.70}{\includegraphics{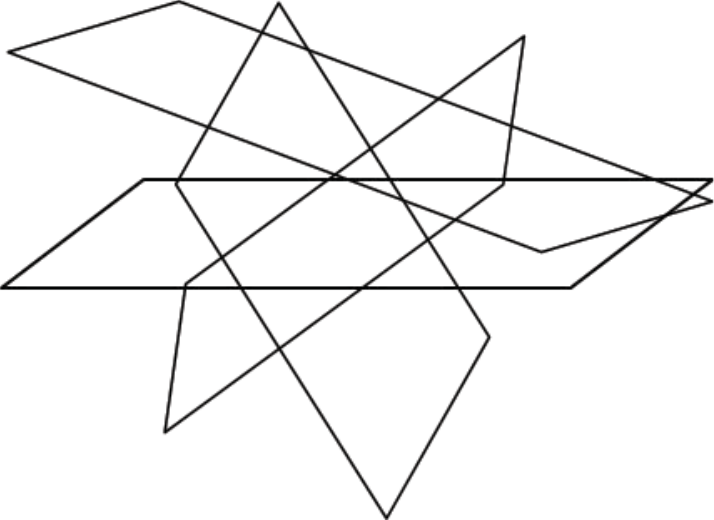}}}
\vspace*{-0.1in}
\caption{Random orientation of lower-dimensional planes may avoid systematic violation of Lorentz invariance.}
\label{random}
\end{figure}

Combining the essence of both extremes, a framework was recently proposed in which the structure of spacetime is fundamentally $(1+1)$-dimensional universe, but is ``wrapped up'' in such a way that it appears higher-dimensional at larger distances \cite{dejan1}.  The structure of space may be envisioned as an $n$-dimensional ordered lattice on which dynamics are confined to (presumably) $n=1-4$, defined by fundamental scales $L_1 < L_2 < L_3 < L_4$.  Physics with $\Lambda < \Lambda_3$ on length scales $L > L_3 = \Lambda_3^{-1}$ will appear three-dimensional.
When the energy (length) scale becomes of order $\Lambda_2 > \Lambda_3 ~(L_2 <  L_3)$, the manifold transitions from $(3+1)$ to $(2+1)$.  If $\Lambda_2 \sim 1~$TeV, planar events and other interesting effects could be observed at the LHC for collisions with $\sqrt{s} \geq \Lambda_2$ \cite{Anchordoqui:2010hi}, in addition to unique signatures of lower-dimensional quantum black hole production \cite{xcgl}. For random orientation of lower-dimensional planes/lines (see e.g. Fig.~\ref{random} ), violations of Lorentz invariance induced by the lattice become non-systematic, and thus evade strong limits put on theories with systematic violation of Lorentz invariance \cite{Mattingly:2005re}. The scales $L_1$, $L_2$, $L_3$ and $L_4$ are the effective lengths on which the lattice becomes $d=1$, $d=2$, $d=3$ and $d=4$ on average. These scales depend on how the fundamental domains are glued into a global conglomerate and may be different from the dimensions of the original fundamental domains. There could be many concrete ways to achieve this, however our discussion does not crucially depend on it.

Beyond this novelty, however, the framework effectively cures all divergences that plague the $(3+1)$-dimensional aspects of current field theory.  For example, the fine-tuning problem is alleviated.  The radiative corrections to the Higgs mass in $d$ space-time dimensions are obtained for some cutoff energy $\Lambda$ from the top, W, and Higgs self-coupling loop diagrams contributions
\beq
\Delta m^2_H \sim \sum_i \int^\Lambda \frac{d^dk}{(2\pi)^d} \frac{1}{k^2 - m_i^2} = F_d(\Lambda)~~.
\label{higgscorr}
\eeq
where the index $i$ refers to the diagram and the function $F_d(\Lambda)$ denotes the total divergence behavior after full evaluation of the contributing Feynman integrals.  While quadratically-divergent for $d=4$ ({\it i.e.} $F_4(\Lambda) \sim \Lambda^2$), the one loop corrections to the Higgs mass in $(2+1)$ dimensions are linearly-divergent in the cutoff scale, while in the $(1+1)$-dimensional case they are only logarithmically divergent.

Furthermore, the problems plaguing $(3+1)$-dimensional quantum gravity quantization programs are solved by virtue of the fact that spacetime is dimensionally-reduced.  Indeed, effective models of quantum gravity are plentiful in $(2+1)$ and even $(1+1)$ dimensions \cite{carlip,Callan:1992rs,Bogojevic:1998ma}.  Similarly, the cosmological constant problem may be explained as a Casimir-type energy between two adjacent ``foliations'' of three-dimensional space as the scale size $L > L_4$ opens up a fourth space dimension. [For related work see \cite{GonzalezMestres:2010pi,Marozzi:2011da,Angel:2010dt,Fiziev:2010je}.]

What makes this proposal of evolving dimensions very attractive is that some evidence of the lower dimensional structure of our space-time at a TeV scale may already exist. Namely,
alignment of the main energy fluxes in a target (transverse) plane has been observed in families of cosmic ray particles \cite{et1986,Mukhamedshin:2005nr,Antoni:2005ce}. The fraction of events with alignment is statistically significant for families with energies higher than TeV and large number of hadrons. This can be interpreted as evidence for coplanar scattering of secondary hadrons produced in the early stages of the atmospheric cascade development.

An interesting side-effect of such a dimensional reduction scheme is the distinct nature of gravity in lower dimensions.  It is well-known that, in a $(2+1)$-dimensional universe, there are no local gravitational degrees of freedom, and hence there are no gravitational waves (or gravitons).  If the universe was indeed $(2+1)$-dimensional at some earlier epoch, it is reasonable to deduce that no primordial gravitational waves (PGWs) of this era exist today.  There is thus a maximum frequency for PGWs, implicitly related to the dimensional transition scale $\Lambda_2$, beyond which no waves can exist.  This indicates that gravitational wave astronomy can be used as a tool for probing the novel ``vanishing dimensions'' framework.

We note the idea of using PGWs and their frequency spectrum to determine dimensional characteristics of spacetime is not new.  It has been shown, for example, that phase shifts can be introduced from PGW interactions with extra dimensions \cite{hogan}.  Similarly, \cite{hsu} demonstrates the thermalization of PGWs via propagation through extra dimensions.  Alternatively, PGWs can reveal the existence of topological Chern-Simons terms in the modified Einstein-Hilbert action \cite{cs}.

In order to determine an approximate value for the cutoff frequency, we revisit the current state of PGW detection.  Standard cosmological theory predicts that gravitational waves will be generated in the pre/post-inflationary regime due to quantum fluctuations of the spacetime manifold. At temperatures below the $2D \rightarrow 3D$  cross over scale, a standard 3D FRW cosmology is assumed, with the usual radiation- and matter-dominated eras.  Gravity waves can be produced at different times $t_* < t_0 = H^{-1}$, when the temperature of the universe was $T_*$.   The co-moving entropy per volume of the universe at temperature $T_*$ can be expressed as a function of the scale factor $a(t)$ as
\beq
S \sim g_S(T)a^3(t)T^3 ,
\eeq
where the factor $g_S$ represents the effective number of degrees of freedom at temperature $T$ in terms of entropy,
\beq
g_{S}(T) = \sum_{i = \rm bosons} g_i \left(\frac{T_i}{T}\right)^3 + \frac{7}{8} \sum_{j= \rm fermions} g_j \left(\frac{T_j}{T}\right)^3
\eeq
The parameters $i,j$ runs over all particle species.  In the standard model, this assumes a constant value for $T \simeq 300~$GeV, with $g_{S}(T) = 106.75$ due to the fact that all species were thermalized to a common temperature.

Assuming that entropy is generally conserved over the evolution of the universe, one can write
\beq
g_S(T_*)a^3(t_*)T_*^3 = g_S(T_0)a^3(t_0)T_0^3
\eeq
with $T_0 = 2.728~$K.

The characteristic frequency of a gravitational wave produced at some time $t_*$ in the past is thus redshifted to its present-day value $f_0 = f_* \frac{a(t_*)}{a(t_0)} $ by the factor \cite{pgw}
\beq
f_0 = \simeq 9.37\times 10^{-5} ~{\rm Hz}~ (H \times {\rm 1~mm})^{1/2} g_*^{-1/12} (g_*/g_{*S})^{1/3} T_{2.728}
\label{freq0}
\eeq
where the original production frequency $f_*$ is bounded by the horizon size of the universe at time $t_*$, {\it i.e.} $f_* \sim \lambda_*^{-1} \sim H_*^{-1}$.  Note that this is an upper bound,and the actual value may be smaller by a factor $\lambda_* \sim \epsilon H_*^{-1}$, although the final result is weakly sensitive to the value $\epsilon \leq 1$ \cite{pgw}.  This quantity can be related to the temperature $T_*$ by noting that, during the radiation-dominated phase, the scale is
\beq 
H^2_* = \frac{8\pi^3 g_* T^4_*}{90 \Mpl^2}
\label{hstar}
\eeq
We note here that we used equations valid only in the $3+1$-dimensional regime. Without the details of an underlying lower dimensional cosmology we do not know the size of a lower dimensional Hubble volume as a function of the temperature (some ambiguities will be discussed later in the text). However, in order to estimate the frequency cut-off, we are approaching the dimensional cross-over from the known $3+1$-dimensional regime. Thus, while Eq.~(\ref{hstar}) is not valid in a lower dimensional regime, it is valid a few Hubble times after the dimensional cross-over. Since most of the 3D volume of the universe comes from the last few Hubble times, this will be a reasonable estimate of the size of the 3D Hubble volume after the dimensional cross-over. If we plug $T_* = 1~$TeV, we see that $H^{-1} \sim {\rm 1~mm}$, which is much larger than TeV$^{-1}$. This is not in contradiction with our assumption that the cross-over happened at $T_* = 1~$TeV since the size of a $2$-dimensional plane/universe could be arbitrarily large before the cross-over. Since the size of the 2D universe does not matter (no gravity waves), the crucial thing here is that the highest frequency that PGWs can carry is limited by the size of a 3D Hubble volume right after the dimensional cross-over, which is given by Eq.~(\ref{hstar})\footnote{This may change if the history of the early universe is radically different from the standard picture. The standard cosmology must kick in at $T \sim $MeV (nucleosynthesis), and the dimensional cross-over happens at $T \sim $TeV, so one has large freedom in formulating an underlying cosmological model of evolving dimensions.}.

\begin{figure}[htb]
\vspace*{-0.1in}
\center{\scalebox{0.2}{\includegraphics{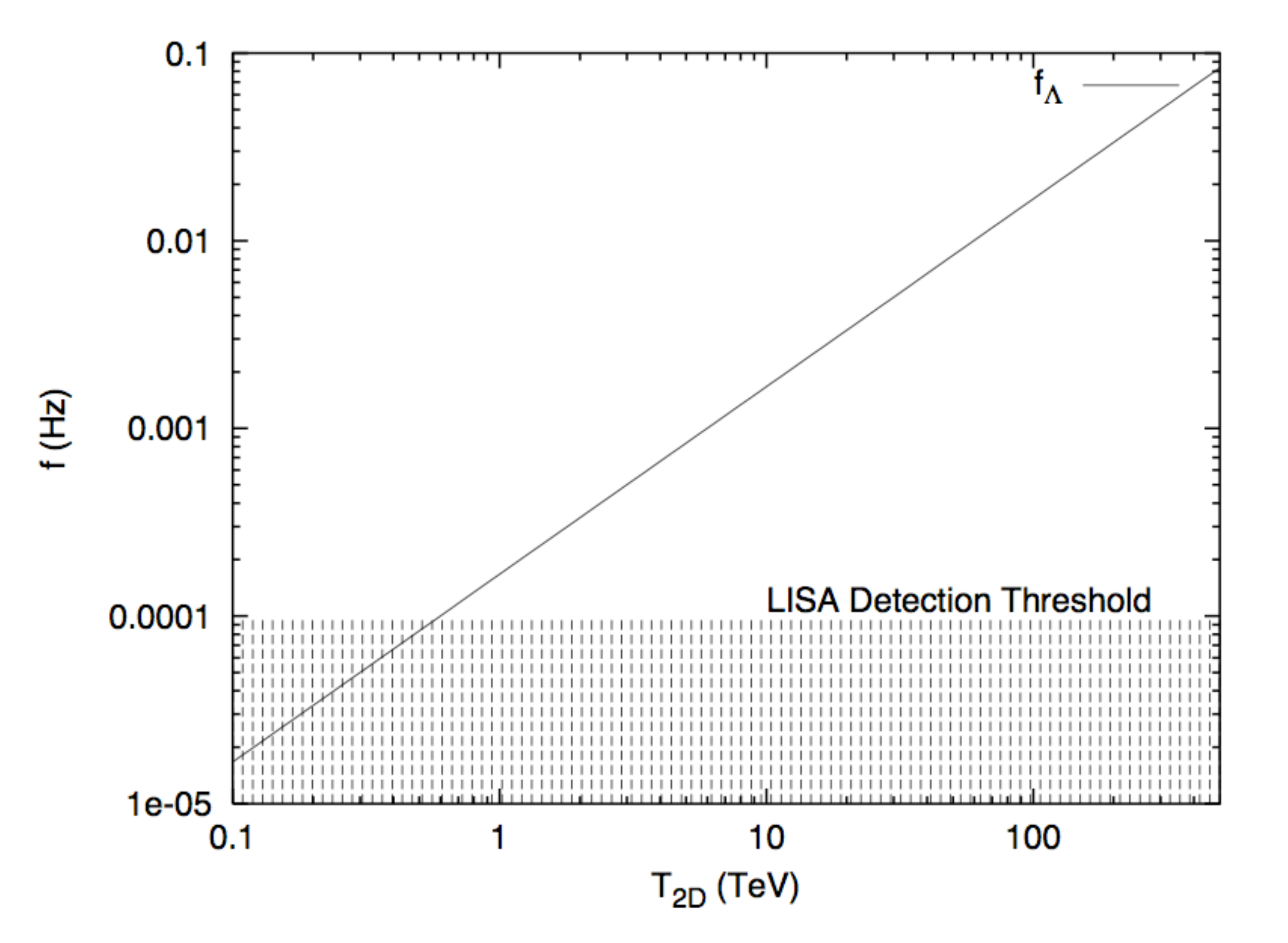}}}
\vspace*{-0.1in}
\caption{Frequency threshold for primordial gravitational waves produced when the universe was at temperature $T_*$.  The hatched region is outside the sensitivity cutoff of LISA. }
\label{freqs}
\end{figure}

With above assumptions, combining Eqs.~\ref{freq0} and \ref{hstar}, the frequency of PGWs that would be detectable is
\bea
f_{\rm \Lambda} & = & 7.655\times 10^{-5}  (g_*)^\frac{1}{6} \left(\frac{T_*}{\rm TeV}\right)~~{\rm Hz}\nonumber \\
 &\approx &1.67 \times 10^{-4} \left(\frac{T_*}{\rm TeV}\right)~~{\rm Hz}
\eea
where the latter equality holds for $g_* \sim 10^2$.  When $T_* = 1~$TeV, the frequency is $f_\Lambda \sim 10^{-4}~$Hz.  This is well below the seismic limit of $f \sim 40~$Hz on ground-based gravity wave interferometer experiments like LIGO or VIRGO \cite{ligo}, but sits precisely at the threshold of LISA's sensitivity range.  Indeed, the latter observatory is expected to probe a variety of early-universe phenomenology from the $100$GeV-$1000$ TeV period \cite{lisa}.  Figure~\ref{freqs} demonstrates the threshold PGW frequency as a function of transition energy $\Lambda_2 = T_{2D}$.

At this point, it is instructive to study the physics of expanding $(1+1)$ and $(2+1)$-dimensional universes in order to check if some unexpected dynamical features can change the conclusions derived so far. We will show below that this does not happen.

In any space-time, the curvature tensor $R_{\mu \nu \rho \sigma }$ may be decomposed into a Ricci scalar $R$, Ricci tensor $R_{\mu \nu }$ and conformally invariant Weyl tensor $C_{\mu \nu \rho }^\sigma $. In three dimensions the Weyl tensor vanishes and  $R_{\mu \nu \rho \sigma }$ can be expressed solely through $R_{\mu \nu }$ and $R$.  Explicitly
\be
R_{\mu \nu \rho \sigma } = \epsilon_{\mu \nu \alpha} \epsilon_{\rho \sigma \beta} G^{\alpha \beta}
\ee
This in turn implies that any solution of the vacuum Einstein's equations is locally flat. Thus, $(2+1)$-dimensional space-time has no local gravitational degrees of freedom, i.e. no gravitational waves in classical theory and no gravitons in quantum theory. Gravity is then uniquely determined by a local distribution of matter. The number of degrees of freedom in such a theory is finite, quantum field theory reduces to quantum mechanics and the problem of non-renormalizability disappears.
A $(2+1)$-dimensional FRW metric is

\beq \label{2dfrw}
ds^2 = dt^2 - a(t)^2 \left(\frac{dr^2}{1-kr^2} + r^2d\theta^2 \right)
\eeq
where $a(t)$ is the scale factor and $k=-1,0,+1$.
The Einstein's equations for this metric are
\beq
\left(\frac{\dot{a}}{a}\right)^2 = 2\pi G \rho -\frac{k}{a^2} \, ,
\ \ \
\frac{d}{dt} (\rho a^2) + p \frac{d}{dt} a^2 =0
\eeq
where $G$ is the $(2+1)$ dimensional gravitational constant, $p$ is the pressure and $\rho $ is the energy density.
In a radiation dominated universe $p=\frac{1}{2} \rho$ and $\rho a^3 =\rho_0 a_0^3 =$const which gives
\beq
\dot{a} =\pm \sqrt{\frac{2\pi G \rho_0 a_0^3}{a}-k} \, ,
\ \ \ \ \
\ddot{a} =\frac{\pi G \rho_0 a_0^3}{a^2}
\eeq
For $k=0$ the solution to these equations is
\beq
a(t) =  \left(\frac{9}{2}\pi G \rho_0 a_0^3\right)^{1/3} t^{2/3}
\label{a21}
\eeq
One can note that three-dimensional solution $a(t) \propto t^{2/3}$ is different from the usual
four-dimensional behavior $a(t) \propto t^{1/2}$ in radiation dominated era.

We have to note here that Einstein's equations, i.e. $G_{\mu \nu} = \kappa T_{\mu \nu}$, are not the only valid option in $(2+1)-$dimensional space. For example it has been known that theories with $R=\kappa T$, where $R$ is the Ricci scalar and $T$ is the trace of $T_{\mu \nu}$, are not good $(3+1)-$dimensional theories of gravity since they do not have a good Newtonian limit. However, in the context of evolving dimensions a good Newtonian limit is not a requirement since the spacetime becomes $(3+1)-$dimensional at distances larger than TeV$^{-1}$. The solutions of $R=\kappa T$ theory were discussed for example in \cite{Cornish:1991kj}. The solution for a radiation dominated universe for the metric given in
(\ref{2dfrw}) is $a(t)=t$.

The crossover from $(2+1)$- to $(3+1)$-dimensional universe happened when the temperature of the universe was $T_{2D \rightarrow 3D} =\Lambda_2 \sim 1~$TeV. Working backwards, we can estimate the size of the Universe at the transition from the ratio of scale sizes at various epochs, specifically between present day ($t_{\rm today} \sim 10^{17}~$s), the radiation/matter-dominated era ($t_{\rm RM} \sim 10^{10}~$s) and the TeV-era ($t_{\rm TeV} \sim 10^{-12}~$s).  The scale factor at the latter epoch is thus
\beq \label{atev}
a_{\rm TeV} = a_{\rm today} \left(\frac{t_{\rm TeV}}{t_{\rm RM}}\right)^{1/2} \left(\frac{t_{\rm RM}}{t_{\rm today}}\right)^{2/3}  = 10^{-15} a_{\rm today}
\eeq
This value may also be obtained by noting that conservation of entropy requires the product $a T$ to be constant, and so $a_{\rm TeV} = 10^{-15} a_{\rm today}$ (since $T_{\rm today} \sim 10^{-3}~$eV). Eq.~\ref{atev} implies that the size of the currently visible universe ($10^
{28}$cm) at $T\sim 1~$TeV was $10^{13}$cm. This distance is macroscopic but it is not in contrast with our assumption that the crossover from $(2+1)$- to $(3+1)$-dimensional universe happened when the temperature of the universe was $T\sim 1~$TeV, since the causally connected universe today contains  many causally connected regions of some earlier time. Finding the exact size of the causally connected universe before the dimensional cross over
is not a unique task, since it would strongly depend on an underlying cosmological model. In particular, it would depend on which scale inflation and reheating happened \cite{Rinaldi:2010yp} (if at all). The absolute lower limit in the energy scale of inflation is about $10$MeV (in order not to affect the earliest landmark of the standard cosmology - nucleosynthesis), but inflation may as well happen at any energy above the dimensional crossover scale.
Further, the dimensional crossover may perhaps be a violent highly non-adiabatic process with huge entropy production. In that case the standard relation $a T=$constant would not be valid anymore at temperatures above TeV (but it would still be valid from $T \sim $TeV till today, as assumed in Eq.~(\ref{atev})).

Fortunately, our limits on PGW are fairly robust. The only explicit input we used was that that highest frequency that PGWs can carry is limited by the size of a 3D Hubble volume right after the dimensional cross-over. The cross-over temperature $T\sim 1~$TeV is the value strongly favored for theoretical reasons \cite{dejan1}, and perhaps also indicated by
the planar events in cosmic ray experiments \cite{et1986,Mukhamedshin:2005nr,Antoni:2005ce}.

Going towards even higher temperatures, the space-time becomes $(1+1)$-dimensional.
To avoid large hierarchy in the standard model, the crossover from an $(1+1)$-dimensional to $(2+1)$-dimensional universe needs to happen when the temperature of the universe was $T_{1D \rightarrow 2D} = \Lambda_1 \leq 100~$TeV.  Conservation of entropy (if between $T\sim 1~$TeV and $T\sim 100~$TeV nothing non-adiabatic happened) requires $aT =$const. This implies
\be
\frac{a_{2D \rightarrow 3D}}{a_{1D \rightarrow 2D}}=\frac{T_{1D \rightarrow 2D}}{T_{2D \rightarrow 3D}} \sim 100
\label{aratio}
\ee
Similarly,
\be
\frac{a_{1D \rightarrow 2D}}{a_0}=\frac{T_{1D \rightarrow 2D}}{T_0}
\ee
where $a_0$ and $T_0$ are the scale factor and temperature of the universe the first time it appears classically. It is tempting to set
$a_0 =M_{Pl}^{-1}$ and $T_0=M_{Pl}$, however $M_{Pl} =10^{19}~$GeV is inherently $(3+1)$-dimensional quantity whose meaning is not quite clear in the context of evolving dimensions.

A $(1+1)$-dimensional FRW metric is \cite{mannchan}
\beq \label{1dfrw}
ds^2 = dt^2 - a(t)^2 \frac{dx^2}{1-kx^2}
\eeq
The denominator in the second term in (\ref{2dfrw}) can be absorbed into a definition
of the spatial coordinate $x$. Moreover, all $(1+1)$-dimensional spaces are conformally flat, i.e. one can always use coordinate transformations (independently of the dynamics) and put the metric in the form $g_{\mu \nu} = e^\phi \eta_{\mu \nu}$. The Einstein's action in a two-dimensional spacetime is just the Euler characteristics of the manifold in question, so the the theory does not have any dynamics, unless the scalar field $\phi$ is promoted into a dynamical field by adding a kinetic term for it.  Even in this case there are no gravitons in theory, so there are no gravity waves and the threshold of importance remains the $2D\rightarrow 3D$ transition.

We finally note that the action for gravity was taken at each step of dimensional reduction to be the dimensionally continued Einstein-Hilbert action. At scales much larger or much smaller than the dimensional crossover this may be justified. However, exactly at the crossover the description could be very complicated. For example, systems whose effective dimensionality
changes with the scale can exhibit fractal behavior, even if they are defined on smooth manifolds. As a good step in that direction, in \cite{Calcagni:2009kc} a field theory which lives in fractal spacetime and is argued to be Lorentz invariant, power-counting renormalizable, and causal was proposed.

Dimensional crossover would also induce non-renormalizable operators suppressed by the crossover energy scale. While their explicit form is unknown, one can in fact put constraints on
their form. Light coming to us from large cosmological distances would be subject to stochastic fluctuations which would induce uncertainties in the wavelength $\delta \lambda \sim L_3 (\lambda/L_3)^ {1-\alpha}$, where $\alpha$ measures the suppression. These are known as non-systematic violations of Lorentz symmetry. Photons that were initially coherent will lose phase coherence as they propagate. For a propagation distance $L$, the cumulative phase dispersion is $\Delta \phi \sim 2 \pi L_3^\alpha L^{1-\alpha}/\lambda$~\cite{Ng:2003ag}.  PKS1413+135, a galaxy at a distance of 1.2~Gpc that shows Airy rings at a wavelength of 1.6~$\mu$m, is a typical probe for effects of this sort. For $\Lambda_3 \sim 1~{\rm TeV}$, the requirement $\Delta \phi \agt 2 \pi$ gives $\alpha \agt 0.8$ \cite{dejan1}, which is not very restrictive.

In conclusion, we proposed a generic and robust test for the new paradigm of ``vanishing" or ``evolving" dimensions where the spacetime we live in is lower dimensional on higher energies.  Since $(2+1)$-dimensional spacetimes have no gravitational degrees of freedom, gravity waves cannot be produced in that epoch. This places a universal maximum frequency at which primordial waves can propagate, marked by the transition between dimensions.  We showed that, under reasonable assumptions, this cut-off frequency may be accessible to future gravitational wave detectors such as LISA. This conclusion may change if the history of the early universe is radically different from the standard picture. Since the standard cosmology must kick in at $T \sim $MeV (nucleosynthesis), and the dimensional cross-over happens at $T \sim $TeV, one has large freedom in formulating an underlying cosmological model of evolving dimensions.

%
 \end{document}